\documentstyle[12pt,nature,psfig]{article}
% ...
\begin{document}
\newcommand{\be}{\begin{eqnarray}}
\newcommand{\ee}{\end{eqnarray}}
\newcommand{\etal}{{\it{et al.}}}
\newcommand{\smass}{M_{\odot}}
\newcommand{\br}{{\bf r}}
\newcommand{\bV}{{{\bf v}}}

\newcommand{\LL}{\mbox{${\log \Lambda}$}}
\newcommand{\Ll}{\mbox{${\log \lambda}$}}
\newcommand{\mstar}{m_\star}
\newcommand{\trxJ}{\mbox{${t_{\rm rJ}}$}}
\newcommand{\trxt}{\mbox{${t_{\rm rt}}$}}
\newcommand{\trxh}{\mbox{${t_{\rm rh}}$}}
\newcommand{\tdf}{{\mbox{$t_{\rm df}$}}}
\newcommand{\trh}{{\mbox{$t_{\rm hr}$}}}
\newcommand{\thc}{{\mbox{$t_{\rm hc}$}}}
\newcommand{\trJ}{{\mbox{$t_{\rm Jr}$}}}
\newcommand{\tcc}{{\mbox{$t_{\rm cc}$}}}
\newcommand{\tdis}{\mbox{${t_{\rm dis}}$}}
\newcommand{\aJ}{${a_{\rm J}}$}
\newcommand{\mJ}{{m_{\rm J}}}
\newcommand{\nJ}{{m_{\rm J}}}
\newcommand{\rJ}{{r_{\rm J}}}
\newcommand{\rcore}{{r_{\rm core}}}
\newcommand{\mcore}{\mbox{${m_{\rm core}}$}}
\newcommand{\rvir}{{r_{\rm vir}}}
\newcommand{\aG}{{a_{\rm G}}}
\newcommand{\MG}{{M_{\rm G}}}
\newcommand{\RG}{{R_{\rm G}}}

\newcommand{\mrunaway}{\mbox{${m_{\rm r}}$}}

\renewcommand{\thefootnote}{\alph{footnote}}

\newcommand{\kms}{\mbox{${\rm km~s}^{-1}$}}
\newcommand{\msun}{\mbox{${\rm M}_\odot$}}
\newcommand{\Msun}{\mbox{${\rm M}_\odot$}}
\newcommand{\Lsun}{\mbox{${\rm L}_\odot$}}
\newcommand{\Rsun}{\mbox{${\rm R}_\odot$}}
\newcommand{\trlx}{\mbox{$t_{\rm rlx}$}}

\def\apgt{\ {\raise-.5ex\hbox{$\buildrel>\over\sim$}}\ }
\def\aplt{\ {\raise-.5ex\hbox{$\buildrel<\over\sim$}}\ }
\def\lt{\ {\raise-.5ex\hbox{$\buildrel>$}}\ }
\def\gt{\ {\raise-.5ex\hbox{$\buildrel<$}}\ }

\def\SN{\ { SN\,2006gy}\ }
\newcommand{\rtide}{\mbox{$r_{\rm tide}$}}

\baselineskip 24pt

%% You can insert a short comment on the title page using the command below.
%\slugcomment{Not to appear in Nonlearned J., 45.}

\def\Ed#1{{\bf[#1 -- Ed]}}
\def\ed#1{{\bf[#1 -- Ed]}}
\def\simon#1{{\bf[#1 -- Simon]}}
\def\Simon#1{{\bf[#1 -- Simon]}}

%
% Title
%
\title{ A runaway collision in a young star cluster as the origin of
the brightest supernova }

%
% Authors
%

\author{
 Simon F. Portegies Zwart\\
 	Astronomical Institute `Anton Pannekoek', \\
 	University of Amsterdam, Kruislaan 403 \\
 {\small and}\\
 	Institute for Computer Science,\\
	University of Amsterdam, Kruislaan 403 \\
 {\small{spz@science.uva.nl}}\\
 ~\\
 Edward P.J van den Heuvel \\
 	Astronomical Institute `Anton Pannekoek', \\
 	University of Amsterdam, Kruislaan 403 \\
{\small and} \\
        Kavli Institute for Theoretical Physics, \\
        University of California Santa
        Barbara, CA 93106-4030, USA \\
 {\small{edvdh@science.uva.nl}}\\
}

%
% Abstract
%

\maketitle

\begin{abstract}
\noindent{\bf 
Supernova 2006gy in the galaxy NGC 1260 is the most luminous one
recorded
\cite{2006CBET..644....1Q,2006CBET..647....1H,2006CBET..648....1P,2006CBET..695....1F}. Its
progenitor might have been a very massive ($>100$\,\msun) star
\cite{2006astro.ph.12617S}, but that is incompatible with hydrogen in
the spectrum of the supernova, because stars $>40$\,\msun\, are
believed to have shed their hydrogen envelopes several hundred
thousand years before the explosion \cite{2005A&A...429..581M}.
Alternatively, the progenitor might have arisen from the merger of two
massive stars \cite{2007ApJ...659L..13O}. Here we show that the
collision frequency of massive stars in a dense and young cluster (of
the kind to be expected near the center of a galaxy) is sufficient to
provide a reasonable chance that SN 2006gy resulted from such a
bombardment. If this is the correct explanation, then we predict that
when the supernova fades (in a year or so) a dense cluster of massive
stars becomes visible at the site of the explosion.
}
\end{abstract}

The presence of hydrogen in supernova SN\,2006gy is hard to reconcile
with the explosion of a $\apgt 40$\Msun\, star, as such a star loses
its hydrogen-rich envelope several hundreds of thousands of years
before the star explodes \cite{2005A&A...429..581M}.  Also the
location of the supernova, at a projected distance of about 1''
($\sim 350$\,pc) from the nucleus of the host galaxy NGC\,1260 is
remarkable.

A merger between a very massive ($>100$\msun) hydrogen-depleted star
that already had a core in an advanced phase of helium burning, with a
hydrogen rich main-sequence star of 10 to 40 \msun, $10^{4}$ to
$10^{5}$ years prior to the supernova explosion may explain the
unusual brightness of the supernova, the presence of the hydrogen in
the interstellar medium surrounding the supernova and the presence of
hydrogen in the supernova itself
\cite{2007ApJ...659L..13O,2006astro.ph.12617S}.

The existence of young star clusters which are in a state of dynamical
core collapse is crucial for the proposed scenario.  During core
collapse and the subsequent post-core collapse evolution of the star
cluster a runaway collision product can grow
\cite{1999A&A...348..117P}, and even though the star is likely to be
much more extended than usual, subsequent bombardment will result in a
net increase in mass \cite{2007astro.ph..3290S}.  Eventually the
massive star is expected to prostrate to a black hole of intermediate
mass \cite{2001ApJ...554..548F,2004Natur.428..724P}.  The supernova in
which the black hole forms is likely to be unusually bright with some
hydrogen in its envelope left-over from the last collision.

The inner few hundred parsec around the center of the Milky Way is
populated with several bright and dense star clusters, of which the
Arches cluster \cite{1996ApJ...461..750C} and Quintuplet
\cite{1992AAS...181.8702C} are the most well known, but many others
exist
\cite{2000A&A...359L...9D,2003A&A...408..127D,2004A&A...423..155M}. The
proximity of the Galactic center and the depth of the potential well
of the bulge causes these clusters to be denser than elsewhere in the
Galaxy \cite{2006ApJ...641..319P}.

The SB/SB0 host galaxy NGC 1260 appears rather ordinary
\cite{1999A&AS..139..141B}, though the presence of a dust lane and of
HII emission near its center suggests that a recent burst of star
formation occurred near its
center \cite{2007ApJ...659L..13O,2006astro.ph.12617S}.  We estimate,
taking an intergalactic extinction of $A \simeq 0.43$\,mag
\cite{2003AJ....126.2268W} into account, that within the observed
isophotal magnitude $B_{25} \simeq 16''$ ($\sim 5.6$\,kpc)
\cite{2003AJ....126.2268W} and adopting $M/L_B \propto L^{0.3}$
\cite{1991ApJ...379...89S}, NGC\,1260 has a mass of $M(5.6{\rm kpc})
\simeq 3 \times 10^{10}$\,\Msun.

Assuming that the mass enclosed within a radius $R$ from the center of
NGC\,1260 is, like in the Milky Way \cite{1972AJ.....77..292S}
described with $M(R) = \mu R^{1.2}$, but for NGC\,1260 $\mu \simeq 9.5
\times 10^5$\,\Msun.  We can then calculate the lower limit to the
tidal radius \cite{1987gady.book.....B} for a cluster of mass $m$ in a
circular orbit at distance $R$ from the center of NGC1260.

Star clusters that experience core collapse before the most massive
stars have left the main sequence can grow a supermassive star via
collision runaway
\cite{1999A&A...348..117P,2002ApJ...576..899P,2004ApJ...604..632G,2006MNRAS.368..141F}.
The mass which can then grow within $\aplt 3$\,Myr can be estimated
using Eq.\,2 of \cite{2006ApJ...641..319P}.  Here we have to make some
assumption about the stellar mass function in the cluster, but for
clarity adopting a mean mass of $\langle m \rangle = 0.5$\,\msun\, is
sufficient without detailed knowledge of the exact shape of the
initial mass function. For a reasonable range of cluster densities and
distances from the center of NGC 1260 we can now calculate the mass
that can be grown in the cluster in $\aplt 3$\,Myr.

In Fig.\,\ref{fig:Mrunaway} we present the results of our calculations
using a King \cite{1966AJ.....71...64K} model with a depth of the
central potential expressed in the dimension-less parameter $W_0=8$,
which can produce at most a $\sim 920$\,\msun\, star in a collision
runaway. For shallower as well as for more concentrated King models
the maximum mass for the supermassive star decreases, as well as the
mass of the cluster that produces such stars.

The last collision before the supernova, must have occurred with a
relatively unevolved main-sequence star, and deposited large
quantities of hydrogen on the surface of the collision product.  By
the time of the supernova not all surface hydrogen of the last
collision was blown away, as about one \msun\, of hydrogen was observed
in the supernova \cite{2006astro.ph.12617S}.  The remainder of the
hydrogen deposited on the stellar surface during the last collision
was found in the interstellar medium surrounding the supernova, and
exceeds some 0.5--5\,\msun\,
\cite{2007ApJ...659L..13O,2006astro.ph.12617S} but could be as high as
20--30\,\msun\, \cite{2006astro.ph.12617S}.  This mass may have come
from the strong stellar wind in the last few $10^4$ years before the
supernova, blowing away the hydrogen which was deposited to the
stellar surface during the collision.  A tentative upper limit for the
rate of mass loss of the progenitor star is $\dot{m} = 5 \times
10^{-4}$\,\msun/yr to $1.4 \times 10^{-4}$\,\msun/yr
\cite{2006astro.ph.12617S}.  These observed mass loss rates are
consistent with those of detailed evolutionary calculations of stars
of 500--$1001$\,\msun\, \cite{2006IAUJD...5E...9Y,2007AASubmitted_Y}.
At this mass loss rate it takes roughly $4 \times 10^{4}$ to $1.4
\times 10^{5}$ years to blow 20\,\msun\, in the form of a stellar wind
from the surface of the supermassive star.  This time scale is of the
same order as our estimated average time between collisions of $\aplt
7.3 \times 10^4$\ years (see fig.\,\ref{fig:Mrunaway}).

The luminosity of the supernova explosion in collapsar models is
driven by the angular momentum transfer from the critically rotating
black hole to its surrounding torus. The available energy reservoir,
and thus the supernova brightness, would then be proportional to the
mass of the black hole \cite{2000PhR...325...83L}. The observed
brightness of SN\,2006gy would then be consistent with the collapse of
an unusually massive star, and the consequent formation of a rather
massive ($\apgt 100$\,\msun) black hole.  We are unaware of detailed
simulations of such an unusual supernova to bolster our arguments, but
the consequences for the supernova seem to be profound and we
encourage further research in this direction.

The amount of hydrogen in the pre-supernova stellar envelope, the
amount of hydrogen in the interstellar medium, the mass-loss rate of
the supernova progenitor derived from the observations and the
enormous brightness of the supernova, bracket the values we derive
based on the collision runaway scenario.  We therefore conclude that a
collision of a $\sim 20$\,\msun\, main-sequence star with a
supermassive star $\sim 10^5$ years before the supernova could
conveniently explain the range of oddities surrounding SN\,2006gy.  We
predict that a young ($\aplt 5$\,Myr), dense and massive ($10^4 \aplt
m \aplt 10^5$\,\msun) star cluster is present at the location of the
supernova. At this moment the star cluster cannot be seen, but
adopting a mass-to-ligh ratio of $\sim 0.6$, which is consistent with
the Starburst99 \cite{1999ApJS..123....3L} models for a $\aplt 5$\,Myr
old stellar population, the cluster should become noticeable as soon
as the supernova fades below an absolute magnitude of about -8.2 mag
for $10^5$\,\msun\, and -5.7\,mag, for a $10^4$\,\msun\, star cluster.

The environment in which the collision runaway can be initiated is
rather exotic, as the cluster has to be sufficiently massive and dense
to warrant dynamical core collapse within a few Myr. In star clusters
sufficiently massive to grow a massive collision product there are
typically between 60 and 600 stars $>8$\,\msun, and consequently only
one out of 60--600 type Ib/c or type II supernovae in these clusters
will be of this peculiar bright type, like SN\,2006gy. If in a nuclear
or a normal starburst ten per cent of all stars are formed in
sufficiently dense clusters, one would expect that about one out of
600-6,000 supernovae to be of this type.

\input ./journals_Nature.def

\section*{Acknowledgments}

This work was supported by NWO under grant No.\, 643.200.503, the
Netherlands Research School for Astronomomy (NOVA) and the National
Science Foundation under grant No.\, PHY 05-51164 to the Kavli
Institute for Theoretical Physics in Santa Barbara.

\begin{figure}
\caption[]{ 
\baselineskip 24pt 
Mass of the collision runaway star as a function of cluster mass and
its distance to the center of NGC\,1260.  
The contours, computed for a King model with $W_0=8$, give the mass of
the runaway collision star as a function of the distance to the center
of the galaxy NGC\,1260 and the mass of the star cluster.  The lowest
four contours are labeled by the mass of the supermassive star (in
solar masses) with constant increments of 100\,\msun\, for subsequent
curves.  A star cluster less massive than about 6,000\,\msun\, is
unable to experience core collapse and produce a collision runaway
before it dissolves in the tidal field of the parent galaxy, whereas
star clusters in the top right corner are unable to reach core
collapse before the most massive stars experience a supernova.
The most massive object that can form is $\sim 920$\,\msun\, in a $m =
1.3 \times 10^5$\,\msun\, cluster.  A $m = 48,000$\,\msun\, cluster
with $W_0 = 5$ can maximally produce a 340\,\msun\, supermassive star,
whereas a King model with $W_0 = 11$ can produce a star of at most
480\,\msun\, in a cluster of 68,000\,\msun.
With an average mass increase per collision of $\sim 20$\,\Msun\,
\cite{1999A&A...348..117P,2002ApJ...576..899P}, the supermassive star
then has experienced at most some 40 collisions between the moment of
gravothermal collapse of the cluster core and the moment that the
supermassive star explodes in a supernova. The mean time between
collisions for this model is then $\aplt 7.3 \times 10^4$\,years.  For
shallower as well as for more concentrated models the time between
collisions is larger; $\aplt 1.7 \times 10^5$\,years for $W_0 = 5$ and
$\aplt 1.2 \times 10^5$\,years for $W_0 = 11$.  For a wide range of
reasonable cluster parameters it appears likely that a collision
runaway ensues and produces a supermassive star.
\label{fig:Mrunaway}
}
\end{figure}

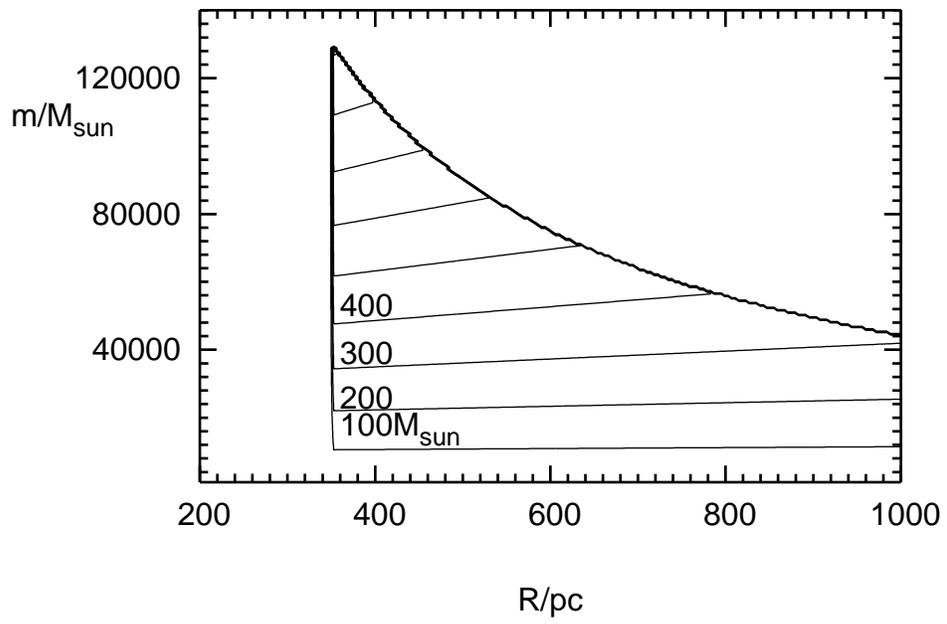
\end{document}